\documentclass[journal]{./IEEEtran}

\usepackage{algorithm}
\usepackage{algpseudocode}
\usepackage{amsfonts}
\usepackage{amsmath}
\usepackage{amssymb}
\usepackage{amsthm}
\usepackage{cite}
\usepackage{color}
\usepackage{graphicx}
\usepackage[breaklinks=true]{hyperref}
\usepackage{mathcom}
\usepackage{multirow}
\usepackage{subfig}

\newtheorem{lemma}{Lemma}

\newtheorem{remark}{Remark}
\newtheorem{theorem}{Theorem}

\algdef{SE}[DOWHILE]{Do}{doWhile}{\algorithmicdo}[1]{\algorithmicwhile\ #1}%

\begin{document}
\title{PaToPa: A Data-Driven Parameter and Topology Joint Estimation Framework in Distribution Grids}
	\author{Jiafan~Yu,~\IEEEmembership{Student~Member,~IEEE}%
	\thanks{J. Yu and R. Rajagopal are with Stanford University, Stanford, CA, 94305 USA. e-mail: \{jfy, ramr\}@stanford.edu},
	Yang~Weng,~\IEEEmembership{Member,~IEEE}%
	\thanks{Y. Weng is with Arizona State University, Tempe, AZ, 85287 USA. e-mail: yang.weng@asu.edu},
	Ram~Rajagopal,~\IEEEmembership{Member,~IEEE}%
	}

\maketitle

\begin{abstract}
The increasing integration of distributed energy resources (DERs) calls for new planning and operational tools.
However, such tools depend on system topology and line parameters, which may be missing or inaccurate in distribution grids. 
With abundant data, one idea is to use linear regression to find line parameters, based on which topology can be identified. 
Unfortunately, the linear regression method is accurate only if there is no noise in both the input measurements (e.g., voltage magnitude and phase angle) and output measurements (e.g., active and reactive power). 
For topology estimation, even with a small error in measurements, the regression-based method is incapable of finding the topology using non-zero line parameters with a proper metric.
To model input and output measurement errors simultaneously, we propose the error-in-variables (EIV) model in a maximum likelihood estimation (MLE) framework for joint line parameter and topology estimation.
While directly solving the problem is NP-hard, we successfully adapt the problem into a generalized low-rank approximation problem via variable transformation and noise decorrelation.
For accurate topology estimation, we let it interact with parameter estimation in a fashion that is similar to expectation-maximization fashion in machine learning.
The proposed PaToPa approach does not require a radial network setting and works for mesh networks.
We demonstrate the superior performance in accuracy for our method on IEEE test cases with actual feeder data from South California Edison.
\end{abstract}

\section{Introduction}
Electric grids are undergoing a revolution towards a more sustainable system.
Renewables and other distributed energy resources (DERs) are expected to supply more than $50\%$ of electricity demand by $2050$~\cite{EU2050, Jacobson2015}. While adding new capabilities, DER penetration raises significant concern about the resilience of power grids, especially in distribution grids.
This is because DERs create two-way power flows, which negatively impact the stability of distribution grids.
Therefore, efficient planning and operational tools are needed for DERs in distribution grids.

Unfortunately, the prerequisite of such planning and operational tools may not be satisfied in current distribution grids. Different from transmission grids, where the topologies and line parameters are regularly measured and verified, such system information in distribution grids can be inaccurate or even unavailable. 
This is because the line parameter profiles in distribution grids are usually only available from grid planning files and the nameplate values, which are likely to be outdated. For example, many substation engineers find it hard to construct the admittance matrix when using distribution management systems, such as CYME~\cite{CYME}. 
Furthermore, in many secondary distribution grids, the topology information is frequently missing. Finally, the measurement and verification tools are limited in distribution grids, making it hard for system operators to track the updates of the line connections, e.g., the real-time switches in distribution grids. 
Without accurate topology and line parameters, real-time monitoring and operational planning are hard to achieve for deep DER integration.
Therefore, distribution grid system operators need new tools to estimate topologies and line parameters.

Fortunately, the ongoing deployment of advanced metering infrastructures (AMIs)~\cite{kwac2014household, yu2015probabilistic} and micro-phasor measurement unit ($\mu$PMU)-type~\cite{sexauer2013phasor} sensors in distribution grids brings abundant data for topology and line parameter estimation.
With the help of GPS timing devices, the existing AMIs could be easily upgraded to have phasor measurement capability in the future.
For example,~\cite{slutsker1996real, liao2010power, indulkar2008estimation, liao2009online, dan2011estimation, prostejovsky2016distribution} use such a data-driven idea.
But, they assume that line measurements are always available, which is usually not widely applicable in distribution grids.
While~\cite{indulkar2008estimation} does not have this drawback, it ignores the noise in the modeling step. 
Furthermore, \cite{davis2013estimation} ignores input measurement error.
While \cite{ding2011transmission, dasgupta2013line} correctly consider both input and output measurement errors for line parameter estimation, their models are only capable of estimating the parameter of a single transmission line with line measurements on both sides, which are incapable of estimating the topology of a distribution grid system.
\cite{yuan2016inverse} renames the joint line and topology estimation as the inverse power flow problem, however, it also does not consider that measurement errors on all variables and only implements a traditional regression model.
Furthermore, even with small error, its regression model is not capable of detecting the topology since almost all elements in the estimated $Y$ admittance matrix are nonzero with small measurement errors.

To accurately model the input-output error and the non-linear power flow equation, we employ the error-in-variables (EIV) model in a maximum likelihood estimation (MLE) problem for parameter estimation~\cite{de1993structured, srebro2003weighted}. 
While the problem is NP-hard, we 1) observe the special structure of the power flow equation, 2) transform variables, 3) linearize the power flow equation at the operating point, and 4) provide matrix relaxation to simplify the objective.
As a result, a generalized low-rank approximation problem is obtained~\cite{van2013total, sato2013riemannian, kukush2005consistency, ottaviani2014exact}.
Furthermore, we propose a method with theoretical performance guarantees, which is based on identity matrix relaxation of the original covariance matrix.
To keep more information from the original covariance matrix, we improve the relaxation from identity matrix to diagonal matrix to maintain the heterogeneity of the accuracy of different measurements.
For the diagonal matrix relaxation, we use an iterative method to find a local minimum solution.
Though there is no theoretical bound for the performance, we found the approach is super robust for different numerical test cases.
In summary, we provide two applicable solutions for the NP-hard problem.
For risk-averse use cases, the identity matrix relaxation could be chosen. 
For use cases that only aims to better numerical results in average, the diagonal matrix relaxation is a good option.

After parameter estimation in a supervised learning framework, we employ an unsupervised learning model to identify the topology.
The proposed topology update method identifies the connected buses and disconnected buses while exchanging information with parameter estimation iteratively. 
The idea is similar to the expectation-maximization algorithm in machine learning, which iterates between maximizing likelihood function in the parameter estimation step and calculating the unknown variables in the expectation step. 
Such an iteration between parameter (Pa) estimation and topology (To) identification is name the PaToPa training flow in our paper.
Finally, for robustness, we also analyze the scenario when there is no angle information for some measurements. 
In such case, we  treat the missing angle information as another source of measurement errors in our EIV model.

We test the proposed approach for joint topology and parameter estimation on different scales of distribution grids, e.g., IEEE 8, 123 bus~\cite{kersting1991radial}, and systems with bus number between 8 and 123.
In all simulations, the estimation results of PaToPa are compared with the results from other state-of-the-art models.
The results reveal that the PaToPa method outperforms other methods for distribution networks for the accuracy for both parameter estimation and topology estimation.
To test the PaToPa method’s robustness, we validate the performance of the model on different level of measurement errors.
The satisfactory results show how to improve monitoring capability with existing data for renewable integrations.

In summary, our contributions are 1) different than the methods in transmission grids, our proposed parameter estimation does not require the topology information, which is a pre-requisite for transmission grid parameter estimation~\cite{abur2004power}, 2) we improve the parameter regression model to tolerate noises in both input and output measurements for various data in distribution grids, 3) we let subsequent topology estimation to actively interact with parameter estimation for joint estimation with better performance, and 4) the solution accuracy has a theoretical guarantee.

The rest of the paper is organized as follows:  Section~\ref{sec:Modeling} provides the problem formulation.  
Section~\ref{sec:EIV} derives the error-in-variables model for line parameter estimation.
Section~\ref{sec:LRA} propose two relaxation methods for solving the line parameter estimation problem.
Section~\ref{sec:PaToPa} introduces the PaToPa training flow for joint line and parameter estimation.
Section~\ref{sec:Exp} numerically demonstrates the superior performance of the proposed method. 
And Section~\ref{sec:Con} concludes the paper.

\section{Problem Modeling for Joint Line Parameter and Topology Estimations}\label{sec:Modeling}
AMIs and smart meters provide real/reactive power injections ($p$, $q$) and voltage magnitude measurements $|v|$.
$\mu$PMU-type measurements can provide voltage phasor information $\theta$.
If there is no noise, $p$, $q$, $|v|$, and $\theta$ can be linked with the admittance matrix through the power flow equation~\cite{lesieutre2011examining}:
\begin{subequations}\label{eqn:PF}
	\begin{align}
		p_i = & \sum_{k=1}^n |v_i||v_k| (G_{ik}\cos \theta_{ik} + B_{ik}\sin \theta_{ik}),\\
		q_i = & \sum_{k=1}^n |v_i||v_k| (G_{ik}\sin \theta_{ik} - B_{ik}\cos \theta_{ik}),
	\end{align}
\end{subequations}
where $i=1, \cdots, n$.
$p_i$ and $q_i$ are the real and reactive power injection at bus $i$, $G$ and $B$ are the real and imaginary parts of the admittance matrix.
$|v_i|$ is the voltage magnitude at bus $i$ and $\theta_{ik}$ is the phase angle difference between bus $i$ and $k$.

For parameter estimation, if we directly estimate $G$ and $B$, there may be overfitting because we ignore the symmetric structure of $G$ and $B$ and the relationships between the $G$ and $B$'s diagonal terms and off-diagonal terms.
To avoid overfitting, we breakdown $G$ and $B$ to estimate the line conductance $\gv$ and susceptance $\bv$ directly.
Since the shunt resistance in distribution grid could be neglected, we can express $G_{ij}$ and $B_{ij}$ as a function of the line parameters $\gv$ and $\bv$, where $g_i$ and $b_i$ are the $i$-th branch's conductance and susceptance, $i = 1, \cdots, m$.
$m$ is the number of possible branches.
For example, if branch $i$ connects bus $j$ and $k$, $G_{jk} = G_{kj} = - g_i$.
Also if $j$-th bus' neighbors are bus $k_1,\cdots, k_l$, and all of its associated branches are branch $i_1, \cdots, i_l$, then the diagonal term $G_{jj} = \sum_{\tau=1}^l g_{i_\tau}$ and $G_{jk_\tau} = - g_{i_\tau}$~\cite{grainger1994power}.
Without loss of generality and to avoid complex notations, we use $v$ to represent $|v|$ afterwards.

With the relationships discussed above, the power flow equations with respect to line parameters are:
\begin{subequations}\label{eqn:PFNew}
\begin{align}	
	&\begin{aligned}
		 p_i =  \sum_{j=1}^m &  g_j |s_{ji}| \left( v_i ^2 - v_{u_{j1}} v_{u_{j2}} \cos (s_{ji}(\theta_{u_{j1}} - \theta_{u_{j2}}))\right)\\
		  & -  b_j |s_{ji}| v_{u_{j1}} v_{u_{j2}} \sin \left(s_{ji}(\theta_{u_{j1}} - \theta_{u_{j2}})\right),
	\end{aligned}\\
	&\begin{aligned}
		 q_i =  \sum_{j=1}^m & b_j |s_{ji}| \left(v_{u_{j1}} v_{u_{j2}} \cos (s_{ji}(\theta_{u_{j1}} - \theta_{u_{j2}})) - v_i ^2\right)\\
		 & - g_j |s_{ji}| v_{u_{j1}} v_{u_{j2}} \sin \left(s_{ji}(\theta_{u_{j1}} - \theta_{u_{j2}})\right),
	\end{aligned}
	\end{align}
\end{subequations}
where $i = 1, \cdots, n$.
$m$ is the number of branches.
$S\in \mathbb{R}^{m\times n}$ is the incidence matrix, e.g., $s_{ij} \in \{1, -1, 0\}$, represents the $i$-th branch leaves, enters or separates from $j$-th bus, respectively.
$U \in \mathbb{R}^{m\times 2}$ is another indexing matrix with $u_{i1}$ and $u_{i2}$ being the ``from'' and ``to'' bus number of the $i$-th branch.

The power flow equation~\eqref{eqn:PFNew} is linear with respect to the line parameters $\gv$ and $\bv$. 
We then transform the variables from the direct measurement $\vv$ and $\thetav$ to a new variable matrix $X$, where 
\[
X = \left[
\begin{array}{cc}
	C & D\\
	D & -C
\end{array}
\right],
\]
and $C, D \in \mathbb{R}^{n \times m}$ with elements
\begin{subequations}\label{eqn:VarTransformation}
	\begin{align}
		c_{ij} = &|s_{ji}| \left( v_i ^2 - v_{u_{j1}} v_{u_{j2}} \cos \left(s_{ji}\left(\theta_{u_{j1}} - \theta_{u_{j2}}\right)\right)\right),\\
		d_{ij} = &- |s_{ji}| v_{u_{j1}} v_{u_{j2}}\sin \left(s_{ji}\left(\theta_{u_{j1}} - \theta_{u_{j2}}\right)\right).
	\end{align}
\end{subequations}
By further assigning the vector $\yv = [\pv; \qv]$ containing the real and reactive power injections, the power flow equations could be written as a mapping from $X$ to $\yv$:
\begin{equation}\label{eqn:PFLinear}
\yv = X \left[\begin{array}{c}\gv\\ \bv\end{array}\right]
\end{equation}

After explicitly defining~\eqref{eqn:PFLinear}, the data-driven joint line parameter and topology estimation problem is:
\begin{itemize}
	\item Given: the historical data of $P = \left(\pv_1, \cdots, \pv_T\right)$, $Q = \left(\qv_1, \cdots, \qv_T\right)$, $V = \left(\vv_1, \cdots, \vv_T\right)$, and $\Theta = \left(\thetav_1, \cdots, \thetav_T\right)$, 
	\item Construct: the new variables $(X_1, \cdots, X_T)$ and $(\yv_1, \cdots, \yv_T)$,
	\item Find: the best estimates of $\widehat{\gv}$ and $\widehat{\bv}$. 
	\item Then, identify the topology based on the estimated line parameters.
\end{itemize}

\section{Line Parameter Estimation through Error-In-Variable Model}\label{sec:EIV}
In this section, we first consider estimating the line parameters for all possible branches and do not consider the estimation of topology.
Without loss of generality, we use $X$ and $\yv$ to compactly denote the ensemble of all the historical data in following context:
\[
X = [X_1; \cdots; X_T],\quad \yv = [\yv_1; \cdots, \yv_T].
\]

In the noiseless case, $X$ and $\yv$ follow the power flow equation~\eqref{eqn:PFLinear} exactly:
\[
\yv = X \left[\begin{array}{c}\gv\\ \bv\end{array}\right],
\]
where $X$ and $\yv$ contain all historical measurements.
Therefore, one can find the line parameters $\gv$ and $\bv$ by solving the linear system in equation~\eqref{eqn:PFLinear}.

\subsection{Measurement Error on Output Only}
However, measurement errors are unavoidable in practice.
To estimate $\gv$ and $\bv$ with noises, a new statistical model is needed.
For example, the maximum likelihood estimation (MLE) criteria can be used for line parameter estimation:
\begin{subequations}\label{eqn:GaussianNonlinear}
	\begin{align}
		\{\widehat{\gv}, \widehat{\bv}\} =  &\underset{\widehat{X}, \widehat{\yv},  \widehat{\gv}, \widehat{\bv}}{\arg \max} \quad P(X, \yv | \{\widehat{X}, \widehat{\yv})\label{eqn:GaussianNonlinearObj},\\
		&\text{subject~to}\quad \widehat{\yv} = \widehat{X} \left[\begin{array}{c}\widehat{\gv}\\ \widehat{\bv}\end{array}\right]\label{eqn:GaussianNonlinearC}.
			\end{align}
\end{subequations}

The estimation problem becomes a least-squares problem if the measurement noise solely comes from the dependent variable $\yv$:
\begin{equation}
\yv = \yv^\star + \epsilonv_{\yv},
\end{equation}
where $\yv^\star$ is the underlining true value satisfying~\eqref{eqn:PFLinear}:
\[
\yv^\star = X \left[\begin{array}{c}\gv\\\bv\end{array}\right].
\]
If the measurement errors of $\yv$ are sampled from i.i.d. Gaussian random variables, the optimization problem has a closed form solution:
\begin{equation}
 \left[\begin{array}{c}\gv^\star_\text{LS}\\ \bv^\star_\text{LS}\end{array}\right] := (X^T X)^{-1} X^T \yv.
\end{equation}

\subsection{Measurement Errors on Both Input/Output: The EIV Model}\label{sec:EIVSub}
However, the assumption that the measurement error only comes from the dependent variable $\yv_t$ is incomplete.
In our case, both power injections $\pv_t, \qv_t$ and voltage phasors $\vv_t, \thetav_t$ are measurements with noises, e.g., PMUs’ calibration error.
Therefore, the indirect measurement $X_t$ also contains induced measurement error, $\epsilon_{X_t}$.
In this case, we have such relationship between the measurements and the true values:
\begin{subequations}
\begin{align}
&\yv = \yv^\star + \epsilonv_{\yv},\\
&X = X^\star + \epsilon_{X},
\end{align}
\end{subequations}
where $X^\star$ and $\yv^\star$ satisfy~\eqref{eqn:PFLinear}:
\[
\yv^\star = {X^\star} \left[\begin{array}{c}\gv\\\bv\end{array}\right].
\]

In this case, the objective~\eqref{eqn:GaussianNonlinearObj} is closely related to the measurement error $\epsilon_X$.
In particular, if the measurement error is i.i.d. Gaussian random variable, the likelihood function could be expressed as a sum of squares.
However, due to the nonlinear variable transformation from the direct measurements $V$ and $\Theta$ to $X$, the noises of $X$ are no longer Gaussian, which makes the MLE problem NP-hard to solve.
In the following, we decorrelate the noise to formulate the problem as a generalized low-rank approximation problem.

If we assume the direct measurement errors of $\vv$ and $\thetav$ are Gaussian and denote the unrevealed ``true'' values of $\vv$, $\thetav$, and $c_{ij}$ at time $t$ as $\vv_t^\star$, $\thetav_t^\star$, and $c_{ijt}^\star$, the measurement error of $c_{ij}$ at time $t$ is a nonlinear function of the measurement errors of $\vv$ and $\thetav$:
\begin{equation}\label{eqn:cijt}
\begin{aligned}
	\epsilon_{c_{ijt}} := & c_{ijt} - c_{ijt}^\star\\
	= & s_{ji} ( v_{it} ^2 - v_{u_{j1}t} v_{u_{j2}t} \cos (\theta_{u_{j1}t} - \theta_{u_{j2}t}))\\
	& - s_{ji} ( {v_{it}^\star}^2 - v_{u_{j1}t}^\star v_{u_{j2}t}^\star \cos (\theta_{u_{j1}t}^\star - \theta_{u_{j2}t}^\star))\\
	= & h(\vv_t, \thetav_t, \epsilonv_{\vv_t}, \epsilon_{\thetav_t})\\
	= & h_{ij}(\epsilonv_{\phiv}; \phiv),
\end{aligned}
\end{equation}
where $\phiv = [\vv; \thetav]$ represents a row vector containing voltage magnitudes and phase angles.
$\epsilonv_{\phiv}$ is the associated direct measurement error.
Similarly, we can express 
\begin{equation}
\epsilon_{d_{ij}} = l_{ij}(\epsilonv_{\phiv}; \phiv).
\end{equation}
Since the noises are usually small quantities, we can use the first order Taylor's expansion for noise approximation:
\begin{equation}
	\begin{aligned}
		\epsilon_{c_{ijt}} = & h_{ij}(\epsilonv_{\phiv_t}; \phiv_t)\\
		= & h(0; \phiv_t) +\epsilonv_{\phiv_t}^T \left.\nabla_{\tauv}h_{ij}(\tauv; \phiv_t)\right|_{\tauv=0}\\
		& + \mathcal{O}\left(\|\epsilonv_{\phiv_t}\|^2\right).\\
	\end{aligned}
\end{equation}
\begin{remark}
The Big-O notation is a relatively loose but convenient expression of which the correctness is numerically proved.
In appendix~\ref{sec:TND}, we use the approximation of the truncated normal distribution to provide a rigorous usage of the first-order approximation.
\end{remark}

By defining
\begin{subequations}\label{eqn:Taylor}
	\begin{align}
	&&\hv_{ij}(\phiv) = &\left.\nabla_{\tauv}h_{ij}\left(\tauv; \phiv\right)\right|_{\tauv=0},\\
	&&\lv_{ij}(\phiv) = &\left.\nabla_{\tauv}l_{ij}\left(\tauv; \phiv\right)\right|_{\tauv=0},
	\end{align}
\end{subequations}
we can get the first order approximation of $\epsilon_{c_{ij}}$ as a function of $\phiv_t$ and $\epsilonv_{\phiv_t}$:
\begin{subequations}\label{eqn:LinearExpression}
	\begin{align}
	&\epsilon_{c_{ij}} = \hv_{ij}(\phiv)^T \epsilonv_{\phiv} + \mathcal{O}(\|\epsilonv_{\phiv}\|^2),\\
	&\epsilon_{d_{ij}} = \lv_{ij}(\phiv)^T \epsilonv_{\phiv} + \mathcal{O}(\|\epsilonv_{\phiv}\|^2).
	\end{align}
\end{subequations}

After the linearization and by ignoring the higher order of error, the measurement errors $\epsilon_{c_{ij}}$ and $\epsilon_{d_{ij}}$ are linear combinations of Gaussian random variables $\epsilonv_{\phiv}$.
Therefore, the elements of $X$ is also Gaussian random variables, and the covariance matrix can be deducted from the covariance matrix of the direct measurement error of $\vv, \thetav$ and the gradients $\hv_{ij}(\cdot)$ and $\lv_{ij}(\cdot)$.
Subsequently, we can express the log likelihood in the objective as a parametric norm associated with the covariance matrix $\Sigma$ of the multivariate Gaussian random variable $\text{vec}([X, \yv] - [\widehat{X},\widehat{\yv}])$:
\[
\log P(\yv, X | \widehat{\yv}, \widehat{X})  = C - \left\|\left[X, \yv\right] - \left[\widehat{X}, \widehat{\yv}\right]\right\|_\Sigma^2,
\]
where $C$ is a constant normalization factor.

The parametric norm is defined below: for any matrix $A \in\mathbb{R}^{m\times n}$, 
\[
\left\|A\right\|_\Sigma = \text{vec}(A)^T \Sigma \text{vec}(A),
\]
where $\Sigma \in \mathbb{R}^{(mn\times mn)}$ is a positive definite matrix, and $\text{vec}(\cdot)$ reshapes a $m\times n$ sized matrix to a $mn \times 1$ sized vector.
Note that $\|\cdot\|_\Sigma$ is a norm because it satisfies the definition of (1) absolute scalability, (2) triangle inequality.

Furthermore, the nonlinear equality constraint~\eqref{eqn:GaussianNonlinearC} could be written as:
\begin{equation}\label{eqn:Xy0}
\left[\widehat{X}, \widehat{\yv}\right]\left[\begin{array}{c}\widehat{\gv}\\ \widehat{\bv} \\ -1\end{array}\right] = 0.
\end{equation}
\eqref{eqn:Xy0} says that there exists an nonzero vector in the null space of $\left[\widehat{X}, \widehat{\yv}\right]$, hence, the matrix $\left[\widehat{X}, \widehat{\yv}\right]$ must be non-full rank matrix.
Therefore, we can transform the nonlinear equality constraint~\eqref{eqn:GaussianNonlinearC} to a matrix low rank constraint:
\begin{equation}\label{eqn:LowRank}
\text{Rank}\left(\left[\widehat{X}, \widehat{\yv}\right]\right) < 2m + 1.
\end{equation}

After having these results, we can express the optimization problem~\eqref{eqn:GaussianNonlinear} as a generalized low-rank approximation (GLRA) problem~\cite{van2013total}:
\begin{subequations}\label{eqn:LRNew}
	\begin{align}
		&\underset{\widehat{X}, \widehat{\yv}}{\min} \quad \left\|\left[X, \yv\right] - \left[\widehat{X}, \widehat{\yv}\right]\right\|_\Sigma^2,\\
		&\text{subject~to} \quad \text{Rank}([\widehat{X}, \widehat{\yv}]) < 2m + 1.
	\end{align}
\end{subequations}

\section{Solving the Generalized Low-Rank Approximation Problem for Parameter Estimation}\label{sec:LRA}
Even if we linearize the objective and express it as a quadratic form associated with the semidefinite matrix $\Sigma$, the generalized optimization problem with arbitrary matrix $\Sigma$ is still very difficult, and the global optimal solution is challenging to obtain~\cite{markovsky2007overview}. 
Therefore, we propose to relax the general $\Sigma$ to more structured shapes based on the patterns of the power flow equation to enable analytical solutions of~\eqref{eqn:LRNew}.

\subsection{Identity Matrix Relaxation for Theoretical Guaranteed Sub-Optimal Solution}
If we ignore the correlation of the induced measurement errors of $X$ and $y$, and assume that their variances are the same, the covariance matrix $\Sigma$ is relaxed to an identity matrix $I$. 
In this case, the original problem~\eqref{eqn:LRNew} becomes a total least square problem:
\begin{subequations}\label{eqn:LRIdentity}
	\begin{align}
		&\underset{\widehat{X}, \widehat{\yv}}{\min} \quad\left\|\left[X, \yv\right] - \left[\widehat{X}, \widehat{\yv}\right]\right\|_F^2,\\
		&\text{subject~to} \quad \text{Rank}([\widehat{X}, \widehat{\yv}]) < 2m + 1,
	\end{align}
\end{subequations}
where $F$ is the matrix frobenius norm.
\eqref{eqn:LRIdentity} has a closed form solution, called Total Least Square (TLS)~\cite{golub1980analysis, van2013total, eckart1936approximation, beck2005global, markovsky2008structured, fuller2009measurement}:
\begin{equation}\label{eqn:TLSSolution}
 	\left[\gv^\star_{\text{TLS}}; \bv^\star_{\text{TLS}}\right] = (X^TX - \sigma_{T+1}^2 I)^{-1}X^T\yv,	
\end{equation}
where $\sigma_{T+1}$ is the smallest singular value of the expanded sample matrix $[X, \yv]$.

In addition, we have the norm equivalence lemma~\cite{bornemann1993basic}:
\begin{lemma}\label{lemma:NormEq}
For any two matrix norms $\|\cdot\|_{\Sigma_1}$ and $\|\cdot\|_{\Sigma_2}$, there exist $c_1$ and $c_2$ independent of $X$, such that for any $X$, 
\[
c_1 \left\|X\right\|_{\Sigma_1} \le \left\|X\right\|_{\Sigma_2} \le c_2\left\|X\right\|_{\Sigma_1}.
\]
\end{lemma}
Therefore, the optimal solution of~\eqref{eqn:LRIdentity} provides a guaranteed bound of the original NP-hard problem~\eqref{eqn:LRNew}:
\begin{theorem}\label{thm:Guarantee}
If $(X^\star_F, \yv^\star_F)$ is an optimal solution of~\eqref{eqn:LRIdentity} and $(X^\star, \yv^\star)$ is an optimal solution of~\eqref{eqn:LRNew}, we have the property of sub-optimality:
\begin{equation}
\left\|\left[X^\star_F, \yv^\star_F\right]\right\|_\Sigma \leq \frac{c_2}{c_1} \left\|\left[X^\star, \yv^\star\right]\right\|_\Sigma.
\end{equation}
\end{theorem}
\begin{proof}
We have
\begin{equation}
\begin{aligned}
&c_1 \left\|\left[X^\star_F, \yv^\star_F\right]\right\|_\Sigma \\
\leq &\left\|\left[X^\star_F, \yv^\star_F\right]\right\|_F\\
\leq &\left\|\left[X^\star, \yv^\star\right]\right\|_F\\
\leq &c_2\left\|\left[X^\star, \yv^\star\right]\right\|_\Sigma.
\end{aligned}
\end{equation}
The first and third inequalities come from Lemma~\ref{lemma:NormEq}, and the second inequality is from the optimality of $X^\star_F$ and $\yv^\star_F$ with respect to the optimization problem~\eqref{eqn:LRIdentity}.
\end{proof}

\subsection{Diagonal Matrix Relaxation for Numerically Enhanced Solution}
Based on Theorem~\ref{thm:Guarantee}, the identity matrix relaxation provides a theoretical guarantee of the optimal solution.
However, the bound is relatively loose with numerical result.
Therefore, we propose another relaxation with better numerical performance.
Furthermore, we can keep the information of the heterogeneity of the diagonal elements by relaxing the original matrix norm $\|\cdot\|_\Sigma$ to a diagonal matrix $\|\cdot\|_{\bar{\Sigma}}$.
In fact, at lease $2nT$ blocks are just scalars in the covariance matrix $\Sigma$.
Therefore, the diagonal relaxation can keep a substantial part of the original structure of the original matrix, which makes it potential to obtain a better result.
In practice, we use an iterative algorithm~\eqref{eqn:DerivativeZeroFinal} to find an analytical solution with numerically verified performance.
To maintain the article self-contained, the algorithm is compactly described in Appendix~\ref{sec:AppWTLS}.

\section{PaToPa: Joint Parameter and Topology Estimation}\label{sec:PaToPa}
Based on the parameter estimation, we propose the PaToPa approach, which consists of iteratively parameter estimation and topology estimation.

In the first ``blind'' parameter estimation step (Pa), all possible connected lines are estimated.
In the second topology estimation step (To), we sort estimated conductances of all possible connected lines and propose a binary search method to opt-out some disconnected lines, since the line conductance ($g$) and susceptance ($b$) are both zero when the two buses are disconnected.

By iteratively updating the topology estimation, we gradually reduce the model complexity for the line parameter estimation which provides a more accurate line parameter estimation.
The detailed topology estimation step is described below.
Given sorted estimated line conductances $\gv$, we first initialize the minimum and maximum search location $i_{\min}$ and $i_{\max}$ to be $0$ and $m-1$, where the length of $\gv$ is $m$.
While the binary search is not finished, we set the current search location $i_{\text{curr}}=(i_{\min} + i_{\max}) / 2$.
We remove all the lines with conductances smaller than the $i_{\text{curr}}$'s conductance, set them as disconnected and retrain the EIV model.
If the associated likelihood of the retrained EIV model is smaller than the likelihood of the model with all lines are assumed to be connected, we mistakenly remove some connected line(s).
Therefore, we update $i_{\max}$ to be $i_{\text{curr}}$.
If the associated likelihood of the retrained model is greater than the likelihood of the model with all lines are assumed to be connected, we safely removed all disconnected lines.
Therefore, we update $i_{\min}$ to be $i_{\text{curr}}$.
The algorithm is shown in Algorithm~\ref{alg:TopoEst}.
To show that Algorithm~\ref{alg:TopoEst} will not mistakenly remove a connected branch, we use Theorem~\ref{thm:Mono} to show why.
\begin{theorem}\label{thm:Mono}
When we set a connected edge to be disconnected, the log-likelihood of the best fit must be smaller than the connected situation when measurement error is small.
\end{theorem}
\begin{proof}
The optimization problem~\eqref{eqn:LRNew} maximizes the log-likelihood
\[
- \left\|\left[X, \yv\right] - \left[\widehat{X}, \widehat{\yv}\right]\right\|_{\Sigma}^2,
\]
where $X$ and $\yv$ are measurements.
By denoting $X = X_0 + \delta X$, $\yv = \yv_0 + \delta\yv$, where $X_0$ and $\yv_0$ are the noiseless values and $\delta X$ and $\delta \yv$ are the measurement noises, we can write the optimal log-likelihood $l$ as a function of the measurement noise $\delta X$ and $\delta \yv$:
\[
l\left(\delta X, \delta \yv\right).
\]
For noiseless $X_0$ and $\yv_0$, there exists true line parameters $\av$ such that $\yv_0 = X_0 \av$, hence the optimal solution of~\eqref{eqn:LRNew} is just $\widehat{X} = X_0$, $\widehat{\yv} = \yv_0$, and the optimal log-likelihood $l(0, 0) = 0$.

We first prove that, if $X_0$, the noiseless value of historical data, is full rank,  there does not exist a $\bv$ with the same dimension as $\av$ satisfying $\yv_0 = X_0^T \bv$ if for some $i$, $b_i=0$, $a_i \neq 0$.
We prove it by contradiction: if there exists such a $\bv$, we have the following:
$\yv_0 = X_0\av = X_0\bv$.
Hence, $X_0(\av - \bv) = 0$.
However, since $X_0$ is full rank, the equality holds only $\av - \bv =0$~\cite{boyd2006linear}, which contradicts to $a_i - b_i \neq 0$.

Actually, if we remove the $i$th column of $X_0$ and $\delta X$ in the optimization problem~\eqref{eqn:LRNew}, we can get the optimal solution of $b$ and the associated log-likelihood:
\[
l_2(\delta X_{-i}, \delta \yv),
\]
where $\delta X_{-i}$ represents the new matrix from removing the $i$th column of $\delta X$.
The noiseless case simply says $l(0, 0) = 0 > l_2(0, 0)$.
The difference is merely determined by the underlying true values $X_0$ and $\yv_0$.
We denote the difference as 
\[
\tau = l(0, 0) - l_2(0, 0).
\]

From~\cite{bonnans1998optimization}, the functions $l(\cdot)$ and $l_2(\cdot)$ are both continuous even if the optimization problem is non-convex. 
Therefore, for any $\tau > 0$, there exists an $\epsilon > 0$, such that for any $\delta X, \delta X_{-i}, \delta \yv$ satisfying $\|\delta X\|_F < \epsilon$, $\|\delta X_{-i}\|_F < \epsilon$, $\|\delta \yv\| < \epsilon$, we have
\[
\|l\left(0, 0\right) - l\left(\delta X, \delta \yv\right)\| < \frac{\tau}{3},
\]
and
\[
\|l_2\left(0, 0\right) - l_2\left(\delta X_{-i}, \delta \yv\right)\| < \frac{\tau}{3}.
\]
Therefore, for small enough measurement errors, the log likelihood of the connected situation is greater than the log likelihood of the case that a connected edge is assumed to be disconnected:
\[
l\left(\delta X, \delta \yv\right) > l_2\left(\delta X_{-i}, \delta \yv\right).
\]
\end{proof}
\begin{algorithm}[htbp]
	\caption{Topology Update}\label{alg:TopoEst}
	\begin{algorithmic}[1]
		\Procedure{TopoEst}{$X, \yv, \gv, \bv, \Sigma$}
				\State $l = \Call{Likelihood}{X, \yv, \gv, \bv, \Sigma}$
				\State $m \gets \text{length of } \gv$
				\State $n \gets \text{length of } \yv$
				\State $i_{\min}, i_{\max} = 0, m-1$
				\While{$i_{\max} > i_{\min} + 1$}
					\State $i \gets \frac{i_{\max} + i_{\min} }{2}$
					\State $\sv \gets \{j\} \subseteq \{0, \cdots, m-1\}$ if $g[j] \ge g[i]$
					\State $\widetilde{X} \gets \left[X[0:2n/3, \sv], X[0:2n/3, \sv + m]\right]$
					\State $\widetilde{\yv} \gets \yv[0:2n/3]$
					\State $\widetilde{X}_v \gets \left[X[2n/3:n-1, \sv], X[2n/3:n-1, \sv + m]\right]$
					\State $\widetilde{\yv}_v \gets \yv[2n/3:n-1]$
					\State $\widetilde{\gv}, \widetilde{\bv} \gets \Call{EIV}{\widetilde{X}, \widetilde{\yv}}$
					\State $\widetilde{l} \gets \Call{Likelihood}{\widetilde{X}_v, \widetilde{\yv}_v, \widetilde{\gv}, \widetilde{\bv}, \Sigma}$
					\If{$\widetilde{l} > l - 0.2|l|$}
					\State $i_{\min} \gets i$
					\Else
					\State $i_{\max} \gets i$
					\EndIf
				\EndWhile	
			\State \textbf{return} $i_{\min}$
		\EndProcedure
	\end{algorithmic}
\end{algorithm}
In Algorithm~\ref{alg:TopoEst}, we use a separate validation set to estimate the likelihood to ensure the efficiency, and set a looser likelihood update criteria: $\tilde{l} > l - 0.2 \|l\| $ of the topology updates.

As the topology estimation result can be used to reversely improve the parameter estimation, we use the Pa step again with the EIV model and the updated topology information. 
Updated parameter estimation will be used for topology estimation iterations until all the branch number matches the expected branch number, e.g., in a tree structured feeder.
The conductance of all disconnected lines are set to zero in last step.
The second and third steps could be iterated more than once to get further topology update.
We stop the iteration when we cannot opt out more disconnected lines.
The algorithm for the proposed PaToPa training flow is shown in Algorithm~\ref{alg:PaToPa}.
\begin{algorithm}[htbp]
	\caption{Parameter and Topology Joint Estimation}\label{alg:PaToPa}
	\begin{algorithmic}[1]
		\Procedure{PaToPa}{$P, Q, V, \Theta, \Sigma_0$}
		\State $\yv\gets \Call{GenY}{P, Q, \mathcal{E}}$
				\State $X\gets \Call{GenX}{V, \Theta, \mathcal{E}}$
				\State $\Sigma \gets \Call{GenVar}{\Sigma_0, \mathcal{E}}$
			\State $\mathcal{E}' \gets \mathcal{K}_n$
			\State $\mathcal{E} \gets \emptyset$
			\While{$\mathcal{E} \neq \mathcal{E}'$}
			\State $\mathcal{E} \gets \mathcal{E}'$
			\State $X, \yv, \Sigma \gets \Call{UpdateData}{X, \yv, \Sigma, \mathcal{E}}$
			\State $\gv, \bv \gets \Call{EIV}{X, \yv, \Sigma}$
			\State $i \gets \Call{TopoEst}{X, \yv, \gv, \bv, \Sigma}$
			\State $\mathcal{E}' \gets \Call{UpdateTopo}{\mathcal{E}, \gv, i}$
			\EndWhile
			\State \textbf{return} $\mathcal{E}, \gv, \bv$
		\EndProcedure
	\end{algorithmic}
\end{algorithm}

\subsection{Recover Admittance Matrix from Joint Topology and Line Parameter Estimation}
After we obtain the estimated topology $\mathcal{E}$ and the associated line parameters $\gv$, $\bv$, we can easily recover the admittance matrix $Y$ with the help of the incidence matrix $S$ and the indexing matrix $U$ introduced in Section~\ref{sec:Modeling}.
\begin{remark}
The proposed PaToPa framework works with mesh networks, in addition to radial networks, since it treats edges equally without using any radial network properties.
\end{remark}

\subsection{Recover Equivalent Admittance Matrix when There are Partial Measurements}
In many distribution grids, the measurements are only available at the root level where the substation/feeder transformers are located, and the leave level where the residential loads and distributed energy resources are located.
In the intermediate level buses, no measurements are available.
When there are no power injections in hidden buses, the distribution grid could be treated as an equivalent network with \emph{only} the root node and leave nodes due to the Kron reduction of the admittance matrix.
In detail, for such a network, we can introduce an equivalent admittance matrix which represents a connected graph among active buses which have non-zero power injections.
For this case, the proposed PaToPa approach can still learn the equivalent topology and line parameters for the reduced equivalent network and could be used for further analysis, such as state estimation and power flow analysis.

\subsection{Parameter Estimation without Strict PMU Requirement}
When the number of available PMU measurements is reduced, we need to evaluate the impact and see if~\eqref{eqn:LRNew} is still computable. 
Due the flexibility of our input error modeling, we can accommodate one or more buses without PMUs with acceptable information loss for parameter estimation.  
The unknown phase angle could be effectively treated as another source of input noises and the measurement error is the angle between zero and the actual value. 
Since this modification only change the observed phase angle, all the following derivation, including the first order Taylor’s expansion could be done in the same manner.

Since the phase angle differences across buses in distribution grids are much smaller than the phase angle differences in transmission grids, a small error will be introduced when we treat the phase angles as zero. 
For example, if PMU measurement $\theta_{u_{i1}}$ is unavailable, it introduces error into $c_{ij}$,  but $c_{ij}$ is still computable via $c_{ij} = |s_{ji}| \left( v_i ^2 - v_{u_{j1}} v_{u_{j2}} \cos \left(s_{ji}\left(0 - \theta_{u_{j2}}\right)\right)\right)$, so $X$ in~\eqref{eqn:LRNew} is still computable. 
After computing the $X$, the associated covariance matrix $\Sigma$ could be derived from the first order Taylor's expansion~\eqref{eqn:Taylor} at a new position $\phiv$, where $\theta_{u_{j1}} = 0$.

In summary, the proposed PaToPa framework is very robust against system complexity and measurements constraints.

\section{Experimental Results}\label{sec:Exp}
\subsection{Numerical Setup}
We test our joint topology and line parameter estimation approach on a variety of settings and real-world data sets.
For example, we use IEEE $8, 16, 32, 64, 96, 123$-bus test feeders.
The IEEE $16, 32, 64, 96$-bus test feeders are extracted from the IEEE $123$-bus system.
The voltage and phase angle data are from two different feeder grids of Southern California Edison (SCE).
The actual voltage and phase angle measurement data from SCE are used to generate the power injection data at each bus on standard IEEE test grids.
Gaussian measurement noises are added to all measurements for the standard IEEE test grids.
The standard deviation of the added measurement error is computed from the standard deviation of the historical data. 
For example, a $10\%$ relative error standard deviation means that the standard deviation of the historical data of some measurement is 10 times the standard deviation of the measurement error.  
The SCE data set's period is from Jan. $1$, $2015$, to Dec. $31$, $2015$.
Simulation results are similar for different test feeders.
For illustration purpose, we use the 8-bus system for performance demonstration.

\subsection{Accuracy of Joint Parameter and Topology Estimation}
\begin{figure}[!t]
	\centering
	\subfloat[Relative Error Standard Deviation $0$.]{\includegraphics[width=0.45\textwidth]{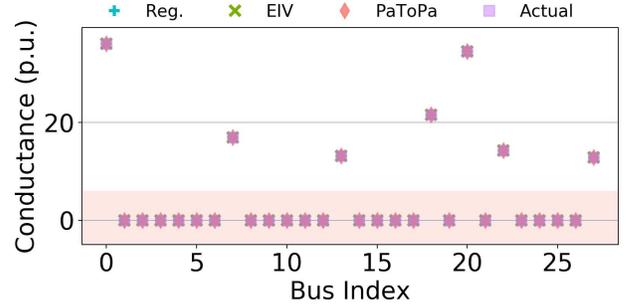}\label{fig:Error0}}\\
	\subfloat[Relative Error Standard Deviation $5\%$.]{\includegraphics[width=0.45\textwidth]{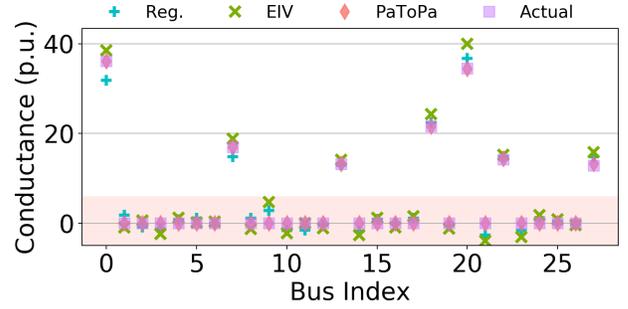}\label{fig:Error5}}
	\\
	\subfloat[Relative Error Standard Deviation $10\%$.]{\includegraphics[width=0.45\textwidth]{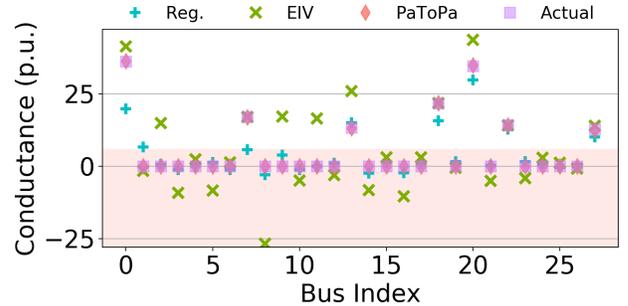}\label{fig:Error10}}
	\caption{
	PaToPa joint topology and parameter estimation illustration. The estimated conductance from different methods and the true values are shown. From top to bottom are the test cases with $0$, $5\%$, and $10\%$ measurement error.
		}
	\label{fig:ErrorVsTopo}
\end{figure}

For an 8-bus system, there are 28 potential connections, represented in the $x$-axis of Fig.~\ref{fig:ErrorVsTopo}.
Since the IEEE 8-bus distribution grid has a radial structure, there are 7 actual connections.
For each potential connection, our joint topology and line parameter estimator will provide a parameter estimation, shown in the $y$-axis.
As there is no noise in the setup that generates Fig.~\ref{fig:Error0}, we observe a perfect match between the estimated line parameters $\widehat{g}$ (red diamonds) and the underlying true value $g^\star$ (purple squares).
In addition to the accurate line parameter estimation, a simple threshold in light red shades can detect the topology perfectly, leading to highly accurate joint topology and line parameter estimation.
When there are noises in both input ($\thetav$, $\vv$) measurements and output ($\pv$, $\qv$) measurements, Fig.~\ref{fig:Error5} and Fig.~\ref{fig:Error10} illustrate the accuracy of the proposed PaToPa method.
Even for $10\%$ relative error standard deviation, when the regression model and EIV model introduce huge error, shown in Fig.~\ref{fig:Error10}, the propose PaToPa model successfully captures these noises, leading to the perfect topology identification and accurate line parameter estimation.

\subsection{Parameter Estimation with Different Noise Levels}
\begin{figure}[!t]
	\centering
	\includegraphics[width=0.45\textwidth]{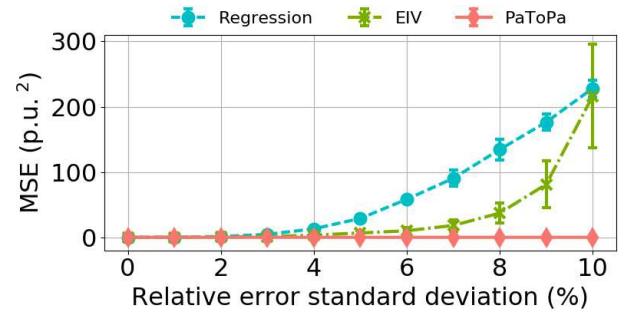}
	\caption{
	The performance of parameter estimation when the measurements are with different levels of noises. The performance is evaluated by the mean squared error of parameter estimation, and the relative noise level is the ratio between the measurement error standard deviation and the historical data standard deviation.
	}
	\label{fig:EIV}
\end{figure}

To summarize different testing results, we look into the statistics in error domain.
As shown in Fig.~\ref{fig:ErrorVsTopo}, when line parameters are estimated accurately, the topology error is small.
Therefore, we focus on the statistical summary of line parameter estimation in Fig.~\ref{fig:EIV}.
For comparison, the performance of the linear regression is plotted in blue, and the performance of the EIV model is plotted in green.
For each error level on the $x$-axis of Fig.~\ref{fig:EIV}, we test the three approaches on 30 different historical data sets with 500 historical data points.
The proposed PaToPa method always has a smaller error than both the regression method and EIV method.
The associated error bars indicate that when there is no measurement error, the performance of both methods are good.
However, with the existence of measurement errors on both input and output, the proposed PaToPa method performs much better than other two methods, showing the robustness of the PaToPa method.

\subsection{Topology Estimation with Different Noise Levels}
Besides the line parameter estimation, we also focus on the accuracies of topology estimation.
To evaluate the accuracy of the topology estimation, we introduce the Jaccard Index, or Jaccard Similarity Coefficient~\cite{jaccard1901etude}, for two sets $A$ and $B$:
\begin{equation}
J(A, B) = \frac{\left|A\cap B\right|}{\left|A\cup B\right|}.
\end{equation}
By assigning $A$ as the set of true connected edges, $B$ as the set of estimated connected edges, we can quantify the accuracy of the topology estimation.
While the regression and EIV models cannot induce the topology estimation directly, we use a threshold method to get the associated topology.
In particular, by setting a conductance threshold, we treat the edges with conductance greater than the threshold as connected, the edges with conductance smaller than the threshold as disconnected.
\begin{figure}[!t]
	\centering
	\includegraphics[width=0.45\textwidth]{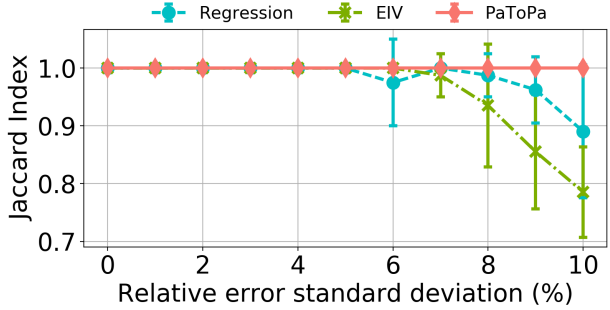}
	\caption{
	The performances of topology estimation when the measurements are with different levels of noise. The performance is evaluated by the Jaccard Similarity Index between the estimated and true connected edges, and the relative noise level is the ratio between the measurement error standard deviation and the historical data standard deviation.
	}
	\label{fig:EIVTopo}
\end{figure}

The results of topology estimation are shown in Fig.~\ref{fig:EIVTopo}.
Consistent with parameter estimation, when the error increases, the accuracy of topology estimation of regression and EIV models degrades.
In particular, the EIV model is more vulnerable to large measurement errors.
This is because of the nonconvexity of the EIV model.
In comparison, the proposed PaToPa method always has zero error no matter how large the error is.

\subsection{Demo Illustration}
\begin{figure}[!t]
	\centering
	\includegraphics[width=0.45\textwidth]{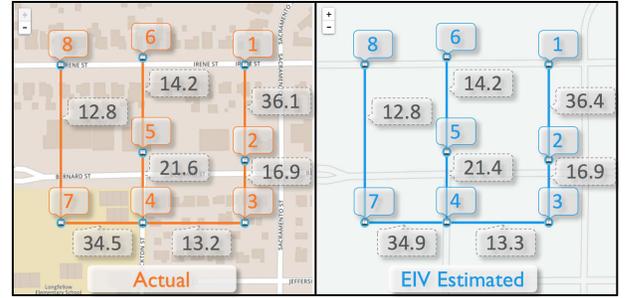}
	\caption{
	Joint estimation for an 8-bus feeder in Bakersfield, CA. The relative standard deviation of  measurement error is $5\%$., and the topology and parameters are estimated through $600$ historical data points. 
	For clearness, the online dashboard only shows the estimated conductance.
	}
	\label{fig:Demo}
\end{figure}

To show the application of the estimation accuracy from above, we test an 8-bus feeder in Bakersfield, CA with a real data set.
We implemented the proposed algorithm on real system based on online map.
Fig.~\ref{fig:Demo} is the result and in production for SCE's system. 
The feeder topology, line parameters, actual input and output data are from SCE's distribution grid SCADA system.
We further add errors to all measurement variables.
Fig.~\ref{fig:Demo} shows a real-time online dashboard with our integrated algorithm running in the back end.
The left panel shows the actual topology and the line conductances in orange.
The right panel shows the estimated topology and line parameters in blue.
Fig.~\ref{fig:Demo} shows that the topology is reconstructed with 0\% error, and the estimated line parameters have a relative error of merely $1\%$.

\subsection{The Choice of Matrix Relaxation Methods For PaToPa}
\begin{figure}[!t]
	\centering
	\includegraphics[width=0.45\textwidth]{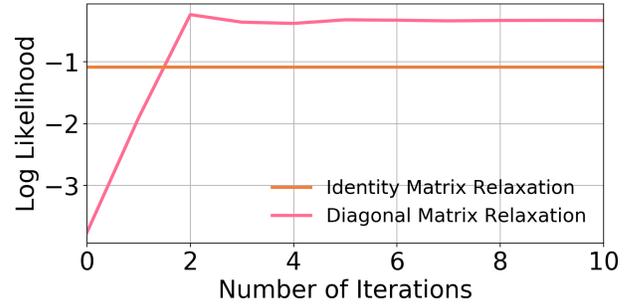}
	\caption{The improvement of log-likelihood over iterations.
	While the identify matrix relaxation results an analytical solution the diagonal matrix relaxation requires the numerical iteration.
	After 2-3 iterations, the log-likelihood of the Diagonal Matrix Relaxation becomes larger than the log likelihood of the identity matrix relaxation. 
	}
	\label{fig:LL}
\end{figure}

We also systematically examine the two matrix relaxation methods in the proposed PaToPa workflow.
In particular, we compare the log likelihood of the identity matrix relaxation with the log likelihoods of the diagonal matrix relaxation during different iterations in computing the results of the generalized total least square problem.
Fig.~\ref{fig:LL} shows the log likelihood of the diagonal matrix relaxation approach for different iterations.
We also plot the log likelihood of the identity matrix relaxation approach as a horizontal line.
After only two iterations, the log likelihood of the diagonal matrix relaxation becomes greater than the log likelihood of the identity matrix relaxation.

\begin{figure}[!t]
	\centering
	\includegraphics[width=0.45\textwidth]{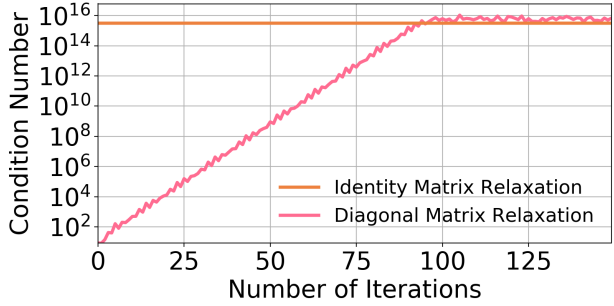}
	\caption{The improvement of the condition number over iterations.
	Only after 100 iterations, the condition number of the optimal solution's matrix with diagonal matrix relaxation is greater than the condition number of the optimal solution's matrix with identity matrix relaxation.
	}
	\label{fig:CN}
\end{figure}

However, we also need to check whether the equality constraint, or the equivalent norm deficiency constraint is satisfied, during the iterations.
The results are shown in Fig.~\ref{fig:CN}.
Notice that the identity matrix relaxation provides the guarantee of the satisfaction of the norm deficiency constraint, which should result infinite condition number.
Due to limited digits in numerical calculation, the condition number for diagonal matrix relaxation solution is on the scale of $10^{15}$.
As we can see, the condition number for diagonal matrix relaxation increases slowly.
The conditional number  achieve the scale of $10^{15}$ after 100 iterations.

Fig.~\ref{fig:LL} and Fig.~\ref{fig:CN} validates the performance of the diagonal matrix relaxation.
Furthermore, we illustrate that the requirement of the number of iterations mainly comes from the constraints, rather than the objective function.

\section{Conclusion}\label{sec:Con}
Detailed system information such as grid topology and line parameters are frequently unavailable or missing in distribution grids. 
This prevents the required monitoring and control capability for deep DER integration.
We propose to extend the existing sensor capability (smart meters, $\mu$PMUs, and inverter sensors) by using their measurement data for joint topology and line parameters estimation in distribution grids. 
Different from other methods which consider only consider output measurement error and assume the topology is already known, our method correctly address the input-output measurement error model with out topology requirement. 
By using the interaction between input-output noises and nonlinear power flow equations in parameter estimation,, we build an error-in-variables model in an MLE problem for joint topology and line parameter estimation. 
With variable transformation and noises decorrelation, we successfully convert the NP-hard problem to a generalized low-rank approximation problem with a closed form solution.
Notably, our proposed approach does not require line measurements, making it robust in distribution grids with a high topology uncertainty. 
Numerical results on IEEE test cases and real datasets from South California Edison show the superior performance of our proposed method.

\section*{Acknowledgment}
The authors thank SLAC National Accelerator Laboratory and South California Edison for their supports.

\appendices
\section{Least Squares and Total Least Squares}
\subsection{Maximum Likelihood Estimation (MLE) for Traditional Regressions}
Traditionally, the parameter estimation model only includes the measurement error in the output.
We generalize the power flow equation as a mapping $f(\cdot; \gv, \bv): \mathbb{R}^n \rightarrow \mathbb{R}$ from the state $\xv$ to the dependent variable $y$.
If both the measurements of $\xv$ and $y$ are noiseless, we have
\begin{equation}
	y = f(\xv; \gv, \bv).
\end{equation}
If the measurement noise solely comes from the dependent variable $y$, the relationship between the measurements $y_t$ and the state $\xv_t$ at time $t$ is
\begin{equation}\label{eqn:LsAssumptionMinus}
y_t - \epsilon_{y_t} = y_t^\star = f(\xv_t; \gv, \bv).
\end{equation}
We usually assume that $\epsilon_{yt}$ is independent of time stamp $t$ and the measurement $y_t$ and is sampled from i.i.d. random variables.
Under this assumption, given a series of measurements $\yv = [y_1, \cdots, y_T]$ and $X = [\xv_1, \cdots, \xv_T]$, we can formulate a Maximum Likelihood Estimation (MLE) problem to find the optimal estimate of the model parameter $\av$:
\begin{equation}
		\gv^\star, \bv^\star = \arg \max_{\gv, \bv}\quad \log P(\yv|X, \gv, \bv).
\end{equation}
We can further write the MLE in a more detailed manner:
\begin{subequations}\label{eqn:MLE}
	\begin{align}
		&\underset{\widehat{\yv}, \gv, \bv}{\max} \quad \log P(\yv | \widehat{\yv}),\\
		&\text{subject~to} \quad \widehat{y}_t = f(\xv_t; \gv, \bv).
	\end{align}
\end{subequations}

Furthermore, if the errors are i.i.d. Gaussian random variables, the log likelihood function is a simple sum of square functions.
In addition, if the function $f$ is a linear function with respect to $[\gv; \bv]$,~\eqref{eqn:MLE} becomes
\begin{subequations}\label{eqn:MLELinear}
	\begin{align}
		&\underset{\widehat{y}, \gv, \bv}{\min} \quad \sum_{t=1}^T (y_t - \widehat{y}_t)^2,\\
		&\text{subject~to} \quad \widehat{y}_t = [\gv; \bv]^T\xv_t.
	\end{align}
\end{subequations}
The solution to such an MLE problem~\eqref{eqn:MLELinear} has a closed-form (Least-Squares) solution:
\begin{equation}
 \left(\gv^\star_\text{LS}, \bv^\star_\text{LS}\right) := (X^T X)^{-1} X^T \yv.
\end{equation}

\subsection{Error-In-Variables: Maximum Likelihood Estimation with Measurement Errors on All Variables}
However, the assumption that the measurement error only comes from the dependent variable $y$ is  incomplete.
In our case, both power injections $\pv, \qv$ and voltage phasors $\vv, \thetav$ are measurements.
Therefore, the noises on all measurements are unavoidable, e.g., the PMUs' calibration error.
Therefore, the mapping relationship~\eqref{eqn:LsAssumptionMinus} turns into:
\begin{equation}\label{eqn:EIVAssumption}
	y_t - \epsilon_{y_t} = f(\xv_t - \epsilonv_{\xv_t}; \gv, \bv).
\end{equation}
Therefore, the MLE problem becomes:
\begin{subequations}\label{eqn:EIV}
	\begin{align}
		&\underset{\widehat{X}, \widehat{\yv}, \gv, \bv}{\max} \quad \log P(X, \yv |\widehat{X}, \widehat{\yv})\label{eqn:EIVObj},\\
		&\text{subject~to} \quad \widehat{y}_t = f(\widehat{\xv}_t; \gv, \bv)\label{eqn:EIVC}.
	\end{align}
\end{subequations}
The difficulty of solving~\eqref{eqn:EIV} comes from the nonlinearity of~\eqref{eqn:EIVAssumption}.

However, when the measurement noise is i.i.d. Gaussian distributed, and the mapping $f$ is linear, a mature technique called low-rank approximation can be used~\cite{van2013total}.
In particular, the log of the probability density function for i.i.d. Gaussian variables is the sum of the measurement error squares.
In this case, the optimization problem~\eqref{eqn:EIV} turns into
\begin{subequations}\label{eqn:EIVGaussian}
	\begin{align}
		&\underset{\widehat{X}, \widehat{\yv}, \gv, \bv}{\min} \quad \sum_{t=1}^T  \|\widehat{\xv}_t - \xv_t\|^2 +  (\widehat{y}_t - y_t)^2,\\
		&\text{subject~to} \quad \widehat{y}_t = (\gv, \bv)^T \widehat{\xv}_t.
	\end{align}
\end{subequations}

Maximizing the log likelihood is equivalent to minimizing the sum of the element-wise squares of the matrix $[X, \yv] - [\widehat{X}, \widehat{\yv}]$, which is the Frobenius norm.
In addition to objective transformation, the constraint could be transformed from $\widehat{y} = [\gv^T, \bv^T] \widehat{\xv}$ to $[\widehat{X}, \widehat{\yv}] [\gv; \bv; -1] = 0$, meaning $[\widehat{X}, \widehat{\yv}]$ is rank deficient and $[\gv; \bv; -1]$ lies in the null space.

Therefore, we can reformulate the parameter estimation problem with measurement errors in both input and output as a low-rank approximation problem for a linear system as:
\begin{subequations}\label{eqn:LR}
	\begin{align}
		& \underset{\widehat{X}, \widehat{\yv}}{\min} \|[X, \yv] - [\widehat{X}, \widehat{\yv}]\|_F^2,\\
		&\text{subject~to} \quad \text{Rank}([\widehat{X}, \widehat{\yv}]) < n + 1.
	\end{align}
\end{subequations}

If the sample matrix $[X, \yv]$ is full rank, \eqref{eqn:LR} has a closed form solution, called Total Least Square (TLS)~\cite{golub1980analysis, van2013total}:
\begin{equation}\label{eqn:TLSSolution}
 	\left[\gv^\star_{\text{TLS}}; \bv^\star_{\text{TLS}}\right] = (X^TX - \sigma_{T+1}^2 I)^{-1}X^T\yv,	
\end{equation}
where $\sigma_{T+1}$ is the smallest singular value of the expanded sample matrix $[X, \yv]$.
However, $f$ is nonlinear for power systems and the resulting MLE problem~\eqref{eqn:EIV} is usually NP-hard at a first look.

\section{The Approximate Behavior of The Truncated Normal Distribution and the Associated MLE Problem}\label{sec:TND}
In Section~\ref{sec:EIVSub}, we first use the first-order approximation of the induced measurement error, and then assume that the induced error random variable is normal.
In this section, we provide a explanation of the detailed approximation procedure.
We revisit the relationship between the original measurement error $\epsilonv_{\phiv}$ and the induced measurement error $\epsilon_c$:
\begin{align}
\epsilon_c = h\left(\epsilonv_{\phiv}\right).
\end{align}
For clearer visualization, without introducing confusions, we remove the subscription $i, j$ for $c$ and $h$, and the parameter $\phiv$ of the function $h$ compared with the main content.

We first assume that the distribution of the direct measurements $\epsilon_{\phiv}$ are sampled truncated normal distribution, truncated at $-d$, $d$ from normal distribution mean $0$, standard deviation $\sigma$.
If we denote the random variable of the truncated normal distribution as $Z(d; \sigma)$, the cumulative distribution function of $Z(d; \sigma)$ is:
\begin{equation}
F(x; \sigma, d) = \frac{\Phi\left(\frac{x}{\sigma}\right) - \Phi\left(\frac{-d}{\sigma}\right)}{\Phi\left(\frac{d}{\sigma}\right)-\Phi\left(\frac{-d}{\sigma}\right)},
\end{equation}
where $\Phi(\cdot)$ is the cumulative distribution function of standard normal distribution.
\begin{theorem}
The truncated normal distribution random variable $Z(d; \sigma)$ converges to the normal distribution with mean zero, standard deviation $\sigma$ in distribution when $d \rightarrow \infty$:
\begin{equation}
\underset{d\rightarrow \infty}{\lim} Z(d; \sigma) \xrightarrow{D} \mathcal{N}\left(0, \sigma\right).
\end{equation}
\end{theorem}
\begin{proof}
Since 
\begin{subequations}
\begin{align}
\underset{d\rightarrow \infty}{\lim} \Phi\left(\frac{d}{\sigma}\right) = 1,\\
\underset{d\rightarrow \infty}{\lim} \Phi\left(\frac{-d}{\sigma}\right) = 0,
\end{align}
\end{subequations}
we have
\begin{equation}
\underset{d\rightarrow \infty}{\lim} F(x; \sigma, d) = \Phi\left(\frac{x}{d}\right).
\end{equation}
\end{proof}

Furthermore, the Taylor's theorem says that
\begin{equation}
\epsilon_c = h(0) + \left.\nabla h(\tauv)\right|_{\tauv=0} \epsilonv_{\phiv} + \frac{1}{2}\epsilonv_{\phiv}^T \left.H_{f\left(\tauv\right)}\right|_{\tauv=\eta \epsilonv_{\phiv}} \epsilonv_{\phiv},
\end{equation}
where $0 \le \eta \le 1$, and $H_f$ is the Hessian of the function $f$.
By introducing this expression, we do not need to consider higher order terms in Taylor's expansion.
If $\epsilonv_{\phiv}$ is a truncated distribution and the truncation range is small enough and the gradient $\nabla h(\tauv)$ is non-zero when $\tauv=0$, the second order term is a higher order error compared with the first order term.
Therefore, by introducing the two-step approximation: Truncated normal distribution and Taylor's expansion, we can use the first-order approximation to express the induced measurement error as a linear function of the direct measurement error.

\section{Diagonal Matrix Relaxation for Generalized Low Rank Approximation Problem}\label{sec:AppWTLS}
If we replace the covariance matrix $\Sigma$ in~\eqref{eqn:LRNew} by a diagonal matrix $\bar{\Sigma}$, and denote $[X, \yv] - [\widehat{X}, \widehat{\yv}] = A$, $[\widehat{X}, \widehat{\yv}] = \widehat{A}$, the objective function could be written as $\sum_i \sum_j w_{ij} (\widehat{a}_{ij} - a_{ij})^2$, where $\bar{\Sigma} = \text{diag}(\{w_{ij}\})$.
Furthermore, the low-rank constraint could be interpreted as $\widehat{A}\cv=0$ for some nonzero $\cv$.
Therefore, the optimization problem for diagonal matrix relaxation could be expressed as:
\begin{subequations}\label{eqn:LRDiag}
\begin{align}
&\underset{\widehat{A}, \cv}{\min} \quad \sum_i \sum_j w_{ij}(a_{ij} - \widehat{a}_{ij})^2\\
&\text{Subject~to:} \quad \widehat{A}\cv = 0,~\cv^T\cv = 1.
\end{align}
\end{subequations}

Then, the Lagrangian of the optimization problem~\eqref{eqn:LRDiag} is:
\begin{equation}
\begin{aligned}
&\mathcal{L}\left(\widehat{A}, \cv, \lv, \lambda \right) \\
= &\frac{1}{2} \sum_i \sum_j w_{ij}(a_{ij} - \widehat{a}_{ij})^2 + \lv^T \widehat{A}\cv + \frac{1}{2}\lambda \left(\cv^T \cv - 1\right)\\
= &0.
\end{aligned}
\end{equation}

By setting the derivatives to zero, we have:
\begin{subequations}\label{eqn:DerivativeZero}
\begin{align}
& w_{ij}(\widehat{a}_{ij} - a_{ij}) = - l_i c_j,\\
& \widehat{A}^T \lv = \cv \lambda,\\
& \widehat{A}\cv = 0,\\
& \cv^T\cv = 1.
\end{align}
\end{subequations}
We further have $\lambda=0$, since $\cv^T\widehat{A}^T\lv = \lambda = 0$.
Without loss of generality, we consider that for all $i$, $j$, $w_{ij} > 0$.
Then we define $V$ as the reciprocal matrix of $W$: $v_{ij} = 1/v_{ij}$.
We further introduce $\dv = \lv / \|\lv\|$, $\sigma = \|\lv\|$.
After defining these,~\eqref{eqn:DerivativeZero} becomes:
\begin{subequations}\label{eqn:DerivativeZeroNew}
\begin{align}
&\widehat{A} = A - \sigma \diag \left(\dv\right) V \diag \left(\cv\right),\label{eqn:DZ1}\\
& \widehat{A}^T \dv = 0,\label{eqn:DZ2}\\
& \widehat{A}\cv = 0,\label{eqn:DZ3}\\
& \cv^T\cv = 1,\\
& \dv^T \dv = 1.
\end{align}
\end{subequations}
By substituting~\eqref{eqn:DZ1} to~\eqref{eqn:DZ2}, we get:
\begin{equation}
	A^T \dv = \sigma \diag\left(\cv\right) V^T \diag \left(\dv\right) \dv.
\end{equation}
By substituting~\eqref{eqn:DZ1} to~\eqref{eqn:DZ3}, we get:
\begin{equation}
	A \cv = \sigma \diag\left(\dv\right) V \diag \left(\cv\right) \cv.
\end{equation}
Then we can define two diagonal matrices:
\begin{equation}
	\begin{aligned}
		&D_d = \text{diag}\left(V^T\diag \left(\dv\right) \dv\right),\\
		&D_c = \text{diag}\left(V\diag \left(\cv\right) \cv\right).\\
	\end{aligned}
\end{equation}
After these preparation, the generalized low-rank approximation problem is converted to:
\begin{subequations}\label{eqn:DerivativeZeroFinal}
\begin{align}
&A\cv = \sigma D_c \dv,\label{eqn:DZF1}\\
& A^T \dv = \sigma D_d \cv,\label{eqn:DZF2}\\
& \cv^T\cv = 1,\\
& \dv^T \dv = 1.
\end{align}
\end{subequations}
An iterative method is proposed to solve~\eqref{eqn:DerivativeZeroFinal}.
We first implement the QR decomposition for $A$:
\[
A = \left[Q_1, Q_2\right] \left[\begin{array}{cc}R\\0\end{array}\right].
\]
With the QR decomposition, we can represent $\dv$ and $\lv$ by two new variable $\zv$ and $\wv$:
\[
\lv = \sigma \dv = Q_1 \zv + Q_2 \wv.
\]
Since $D_c \dv \in \mathcal{R}(A)$ from~\eqref{eqn:DZF1}, we have $Q_2^TD_c \dv = 0$.
Therefore, we get the update rule for $\wv$ from $\zv$:
\begin{align*}
& \sigma Q_2^T D_c \dv = 0 =  Q_2^T D_c Q_1 \zv + Q_2^T D_c Q_2 \wv\\
\Rightarrow & \wv^{[k]} = - \left(Q_2^T D_c^{[k-1]} Q_2\right)^{-1}\left(Q_2^T D_c^{[k-1]} Q_1\right) \zv^{[k]}.
\end{align*}
Then we can update $\lv$ from the definition:
\begin{align*}
&\lv^{[k]} = Q_1 \zv^{[k]} + Q_2 \wv^{[k]},\\
&\xv^{[k]} = \frac{\lv^{[k]}}{\left\|\lv^{[k]}\right\|}.
\end{align*}
From~\eqref{eqn:DZF1} we get the update rule for $\cv$ and $\sigma$:
\begin{align*}
&\cv^{[k]} = \frac{R^{-T}Q_1^{-1}D_c^{[k-1]}\dv^{[k]}}{\left\|R^{-T}Q_1^{-1}D_c^{[k-1]}\dv^{[k]}\right\|},\\
&\sigma^{[k]} = \frac{\left\|c^{[k]}\right\|}{\left\|R^{-T}Q_1^{-1}D_c^{[k-1]}\dv^{[k]}\right\|}.
\end{align*}
Then we can compute $D_d^{[k]}$ and $D_c^{[k]}$ from $\dv^{[k]}$ and $\cv^{[k]}$.
The algorithm is shown in Algorithm~\ref{alg:WTLS}~\cite{de1993structured}:
\begin{algorithm}[htbp]
	\caption{GLRA for Diagonal Matrix Relaxation}\label{alg:WTLS}
	\begin{algorithmic}[1]
		\Procedure{ParamEst}{$A, \bar{\Sigma}$}
		\State $W \gets \text{matrix}\left(\text{diag}(\bar{\Sigma})\right)$
		\State $V \gets \text{reciprocal}\left(W\right)$
		\State Initialize $\cv$, $\dv$
			\While{Not Converging}
			\State $D_d \gets \Call{GetDd}{\dv, V}$
			\State $D_c \gets \Call{GetDc}{\cv, V}$
			\State $\zv \gets R^{-T}D_d \cv$
			\State $\wv \gets - \left(Q_2^T D_c^{[k]} Q_2\right)^{-1}\left(Q_2^T D_c^{[k]} Q_1\right) \zv$
			\State $\dv \gets Q_1 \zv + Q_2 \wv$
			\State $\dv \gets \frac{\dv}{\left\|\dv\right\|}$
			\State $\cv \gets R^{-T}Q_1^{-1}D_c \dv$
			\State $\sigma \gets \frac{1}{\left\|\cv\right\|}$
			\State $\cv \gets \sigma \cv$
			\EndWhile
			\State $\widehat{A} \gets A - \sigma \dv^T V \cv$
			\State \textbf{return} $\widehat{A}$
		\EndProcedure
	\end{algorithmic}
\end{algorithm}

\bibliographystyle{IEEEtran}
\bibliography{bibTex}

\end{document}